# Cluster Rearrangement by Chiral Charge Order in Lacunar Spinel GaNb$_4$Se$_8$

Shunsuke Kitou,[1,*] Masaki Gen,[2] Yuiga Nakamura,[3] Yusuke Tokunaga,[1] Taka-hisa Arima[1,2]

[1]*Department of Advanced Materials Science, The University of Tokyo, Kashiwa 277-8561, Japan.*

[2]*RIKEN Center for Emergent Matter Science, Wako 351-0198, Japan.*

[3]*Japan Synchrotron Radiation Research Institute (JASRI), SPring-8; Hyogo 679-5198, Japan.*

ABSTRACT: Transition-metal atoms with $d$ electrons sometimes form clusters in crystals, which significantly affect the physical properties. Such a cluster formation frequently accompanies a change in the crystal system, leading to the presence of domains with different crystal orientations. In particular, the cubic symmetry is rarely retained after the cluster formation. Here, we identify a cubic-to-cubic phase transition in lacunar spinel GaNb$_4$Se$_8$, where the change in the lattice parameter is less than 0.0001%. Each Nb$^{3.25+}$ tetramer with seven $4d$ electrons is distorted into an Nb$^{3+}$ trimer and an Nb$^{4+}$ monomer induced by charge disproportionation among Nb ions. While the Nb$^{3+}$ trimer with six $4d$ electrons forms spin-singlets in the $\sigma$-bonding orbitals for three Nb-Nb bonds, a localized $S$ = 1/2 spin remains on the $Nb^{4+}$ ion. Furthermore, a local electric dipole moment is induced along the 3-fold rotation axis of each distorted tetramer by the cluster rearrangement. The electric dipole moments are regularly arranged to maintain cubic symmetry, giving rise to chiral order.

INTRODUCTION

Transport and magnetic properties in transition-metal chalcogenides are dominated by the orbital state of transition metals surrounded by the chalcogen elements of O, S, and Se. Early transition-metal atoms sometimes form clusters,[1,2] lowering the energy by sharing bonding $d$ electrons. For instance, rutile-type vanadium dioxide $VO_2$ with a tetragonal lattice undergoes a metal-insulator transition due to V-V dimerization,[3-7] leading to monoclinic distortion of the entire lattice. Such significant changes in physical properties due to cluster formations may be phase-controlled by external fields, allowing for applications in materials such as memory[8,9] and sensors.[10,11] However, cluster formations frequently induce substantial local distortion, causing pronounced anisotropic deformation of the overall lattice and the emergence of domains with varying crystal orientations. In particular, the cubic symmetry is rarely retained after the cluster formation. The significant distortion becomes the primary rate-limiting factor for phase switching. Therefore, the search for materials that form clusters without a reduction in lattice symmetry and with minimal lattice distortion is crucial not only for fundamental research but also for the development of practical materials.

$\beta$-Pyrochlore oxide $CsW_2O_6$ is a rare material that preserves cubic lattice symmetry across the phase transition accompanied by cluster formations. $CsW_2O_6$ with a half-integer average valence of $W^{5.5+}$ ($5d^{0.5}$) undergoes a metal-insulator transition at 215 K,[12,13] where the space group changes from $Fd\overline{3}m$ to $P2_13$. The pyrochlore network of $W^{5.5+}$ in the higher-temperature cubic phase is distorted three-dimensionally to form regular $W^{5.33+}$ ($5d^{2/3}$) trimers and isolated $W^{6+}$ ($5d^0$) ions in the lower-temperature cubic phase, where the lattice constant increases by ~0.004% with the cubic-to-cubic phase transition. The $W^{5.33+}$ trimer has two $5d$ electrons, indicating a three-centered-two-electron bond formation.[13] Such a cluster formation leads to the nonmagnetic insulating nature in the lower-temperature cubic phase. In this system, the cubic symmetry is maintained by orienting adjacent W trimers not parallel to each other.

Lacunar spinel compounds $GaM_4X_8$ ($M$ = V, Nb, Mo, Ta, and $X$ = S, Se) are Mott insulator systems that exhibit various electronic and magnetic properties, such as pressure-induced superconductivity,[14-17] negative colossal magnetoresistance,[18] spin-singlet state,[19,20] spin-orbit entangled molecular $J_{eff}$ state,[20-25] and magnetic skyrmions.[26-29] Since the formal valence of $M$ is 3.25+, the average number of $d$ electrons are 7/4 and 11/4 for group V ($M$ = V, Nb, Ta) and group VI ($M$ = Mo) transition metals, respectively. Since only half of the tetrahedral sites are occupied by Ga atoms, the pyrochlore network of $M$ atoms takes on a breathing-type structure, resulting in $M_4$ tetrahedral clusters. The space group is noncentrosymmetric $F\overline{4}3m$. As cooling $GaM_4X_8$ with $M$ = V and Mo, the overall lattice is distorted along one of the cubic <111> axes, resulting in the rhombohedral polar $R3m$ space group.[30,31] This rhombohedral distortion is understood as a cooperative Jahn-Teller distortion of the $M_4$ molecular orbitals. Furthermore, in a magnetic field, magnetic orders including a skyrmion lattice emerge at lower temperatures in the rhombohedral phase.[26-29] In contrast, Ga$M_4X_8$ with $M$ = Nb and Ta exhibit different types of distortion. The lattice of GaTa$_4$Se$_8$ changes to tetragonal below $T_M$ = 50 K.[20,25] GaNb$_4$S$_8$ and GaNb$_4$Se$_8$ have orthorhombic distortion in the low-temperature phase below $T_M$ = 32 and 33 K, respectively.[20,32] These low-temperature phases are characterized by nonmagnetic (spin-singlet) nature confirmed by nuclear magnetic resonance.[19,20] However, GaNb$_4$Se$_8$ undergoes another phase transition at $T_C$ = 50 K.[20] Powder X-ray diffraction (XRD) experiments suggest the cubic space group $P2_13$ in the intermediate-temperature phase between $T_M$ and $T_C$.[20]

In this study, we perform single-crystal synchrotron XRD experiments to reveal the crystal structure in the intermediate-temperature phase of GaNb$_4$Se$_8$. Nb clusters are found to be rearranged to form a chiral charge order associated with charge disproportionation among Nb ions. The cubic-to-cubic phase transition maintains the overall lattice and total magnetic moment with an exquisite cluster rearrangement, which is different from the first-order phase transition to a nonmagnetic state in CsW$_2$O$_6$.[13]

EXPERIMENTS

Polycrystalline samples of GaNb$_4$Se$_8$ were synthesized by a solid-state reaction of high-purity ingredients Ga (99.99%), Nb (99.9%), and Se (99.999%). The stoichiometric powder, weighing 2 g, was sealed in an evacuated quartz ampule and heated at 900 °C for 70 h. 0.5 g of the obtained polycrystalline material was used as a source for single-crystal growth by chemical vapor transport. The growth was performed using PtCl$_2$ as a transport agent in a quartz ampule at temperatures between 900 and 950 °C for 240 h.

Magnetization measurements were performed by a superconducting quantum interference device (MPMS, Quantum Design) between 2 and 300 K in magnetic fields of up to 7 T. Heat capacity was measured by the thermal relaxation method by using a commercial system (Quantum Design: PPMS). Thermal expansion $\Delta L/L$ along the [111] axis was measured by the fiber-Bragg-grating (FBG) technique using an optical sensing instrument (Hyperion si155, LUNA) in a cryostat equipped with a superconducting magnet (Oxford Spectromag). Although the FBG method causes quantitative errors due to incomplete coupling between the fiber and the sample,[33] one can detect the sample strain with an accuracy of $\Delta L/L \sim 0.0001\%$. XRD experiments were performed using a single crystal of $52 \times 48 \times 35$ $\mu m^3$ on BL02B1 at a synchrotron facility SPring-8 in Japan.[34] A He-gas-blowing device was employed for controlling the temperature between 34 and 300 K. The X-ray wavelength was $\lambda = 0.30956$ Å. A two-dimensional detector CdTe PILATUS was used to record the diffraction pattern. The intensities of Bragg reflections with the interplane distance $d > 0.28$ Å were collected by CrysAlisPro program.[35] Intensities of equivalent reflections were averaged and the structural parameters were refined by using Jana2006.[36] Crystal structures are visualized by using VESTA.[37]

RESULTS AND DISCUSSION

Figures 1a and 1b show the crystal structure of $GaNb_4Se_8$ at 70 K in the high-temperature phase. Merohedral domains corresponding to space inversion were not observed. The deficiency of the Ga site is found to be less than 1%. Nb atoms form regular tetrahedra. The intra- and inter-clusters distances between Nb atoms are 3.0345(4) and 4.3295(5) Å, respectively. Details of the structural parameters are summarized in Tables S1 and S2. Figure 1c shows the molecular orbital scheme of a $Nb_4$ cluster. The $t_{2g}$ atomic orbitals of a Nb ion in a regular $Se_6$ octahedron split into five types of molecular orbitals with different energies under $T_d$ symmetry due to the $Nb_4$ cluster formation.[38] The $t_2$ molecular orbitals accommodate one electron, resulting in an $S$ = 1/2 state when ignoring the spin-orbit interaction.

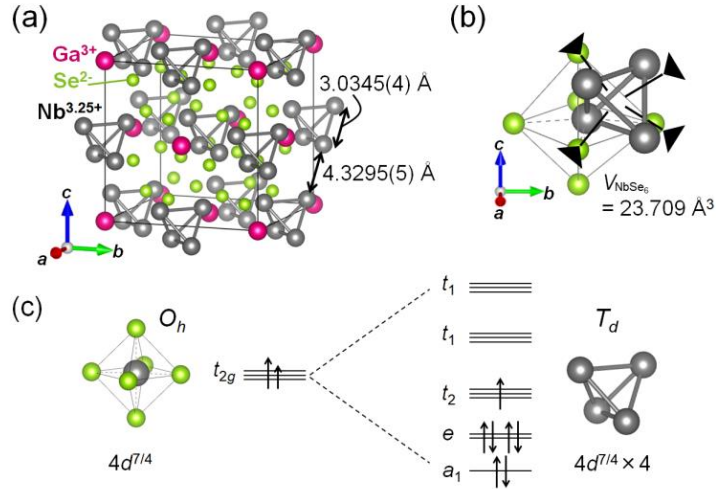

**Figure 1.** (a) Crystal structure of $GaNb_4Se_8$ at 70 K in the high-temperature phase. (b) $Nb_4$ tetrahedron (gray) and $Se_6$ octahedron (green) in $GaNb_4Se_8$. Solid triangles indicate 3-fold rotation axes along <111>. (c) Schematic of the molecular orbital formation of the Nb tetramer. The $t_{2g}$ atomic orbitals under $O_h$ symmetry at the four Nb sites form five molecular orbitals with different energies under $T_d$ symmetry.

Figure 2a shows the temperature dependence of magnetic susceptibility $\chi$ for H || [111] of $GaNb_4Se_8$. There is no anisotropy in the magnetic susceptibility when the external magnetic fields are applied in the [111], [110], and [001] directions (Figure S1 and Table S5). We perform a Curie-Weiss fit for the magnetic susceptibility curve using a formula

$\chi(T) = \chi_0 + C/(T - \Theta)$, where $\chi_0$, $C$, and $\Theta$ are temperature independent components of $\chi$, Curie constant, and Weiss temperature, respectively. The fit above 150 K yields $\chi_0 = -6.2 \times 10^{-5}\,\text{emu/mol}$, $C = 0.3033\,\text{K emu/mol}$, and $\Theta = -142.6\,\text{K}$. The Curie constant gives the effective magnetic moment $\mu_{\text{eff}} = 1.558\mu_B/\text{f.u.}$, which is slightly smaller than the expected value of $1.73\mu_B/\text{f.u.}$ for $S$ = 1/2. A kink and a sharp decrease are observed in $\chi$ at $T_C$ = 50 K and $T_M$ = 31 K, respectively. These results are consistent with previous reports.[20,39] A thermal hysteresis indicates the first-order nature of the phase transition around $T_M$, as shown in the inset of Figure 2a. Furthermore, $T_M$ slightly decreases with increasing external magnetic field (Figure S2).

Figure 2b shows the temperature dependence of heat capacity divided by temperature ($C_p/T$) at 0 T. Broad and sharp peaks are observed at $T_C$ and $T_M$, which indicate second- and first-order like phase transitions, respectively. The low-temperature behavior deviates from the Debye model ($C_p \propto T^3$) (Figure S3a). The $C_p \propto T^2$ component appears at low temperatures, which is not a typical behavior of a three-dimensional antiferromagnet, where the magnon contribution $C_p^{\text{mag}}$ is proportional to $T^3$. The entropy change $\Delta S_C$ = 1.0 J/(mol K) associated with the second-order like phase transition at $T_C$ is estimated from $C_p$ in 40 K ≤ $T$ ≤ 50 K. Since the heat capacity was measured by the thermal relaxation method, the released entropy in the first-order phase transition at $T_M$ could not be accurately estimated from $C_p$. The entropy change at $T_M$ is estimated from the temperature dependence and the external magnetic field dependence of the magnetic susceptibilities as $\Delta S_M$=6.8 J/(mol K) (Figure S2). In the heat capacity measurements, $T_M$ decreases with increasing the external magnetic field (Figure S3c), which is consistent with our magnetic susceptibility results and the previous report.[39] In contrast, $T_C$ slightly increases with increasing external magnetic field (Figure S3b).

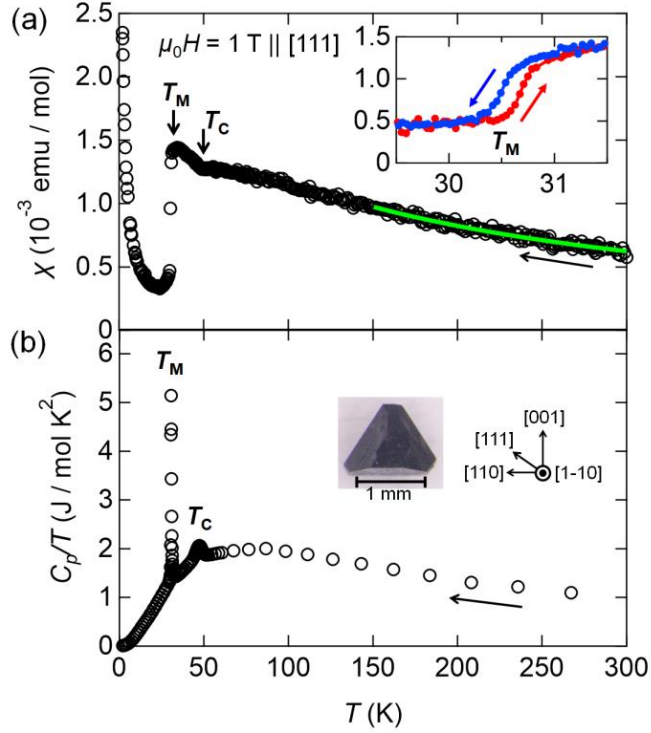

**Figure 2.** (a) Temperature dependence of the magnetic susceptibility $\chi$ for H ∥ [111] of GaNb$_4$Se$_8$. The yellow-green line indicates the Curie-Weiss fit above 150 K. The inset shows an enlarged view of the behavior around $T_M$. Blue and red circles indicate the data in the cooling and heating processes, respectively. (b) Temperature dependence of the heat capacity divided by temperature, $C_p/T$, at 0 T. The inset shows a single crystal used for the magnetic susceptibility and heat capacity measurements.

Figure 3a shows the temperature dependence of thermal expansion $\Delta L/L_{300K}$ along the [111] axis of GaNb$_4$Se$_8$. While a clear increase of about 0.0002% in $\Delta L/L_{300K}$ is observed at $T_M$, no anomaly is observed around $T_C$ with a change of less than 0.0001%. Figure 3b shows the temperature dependence of the integrated intensity of a Bragg peak −5 2 0, which appears below $T_C$. The appearance of the Bragg reflection of −5 2 0 is attributable to the change from the $F$ to the $P$ lattice. Figures 3c and 3d show the XRD data on the $H K 0$ plane at 70 and 40 K, respectively. Although new Bragg peaks appear at reciprocal lattice points with an odd $h$ or odd $k$, no peak splitting is observed for Bragg reflections in the intermediate-temperature phase. The systematic absence of Bragg peaks at $h00$: $h = 2n+1$ and at $0k0$: $k = 2n+1$ (red triangles in Figure 3d) shows $2_1$ screw axes along the principal axes in the intermediate-temperature phase.

Figure 4a shows the crystal structure of GaNb$_4$Se$_8$ at 40 K in the intermediate-temperature phase. The space group is $P2_13$, which is consistent with the former powder XRD experiment.[20] Details of the structural parameters are summarized in Tables S3 and S4. There are two distinct Nb sites, Nb1 and Nb2, with a ratio of 3:1. There is a difference in the volumes between Se$_6$ octahedra around Nb1 (gray) and Nb2 (orange), as shown in Figure 4b, which is ascribed to the difference in the valence of the Nb ions. Since the ionic radius of Nb$^{3+}$ and Nb$^{4+}$ are 0.72 and 0.68 Å,[40] Nb ions coordinated by the larger and smaller octahedra correspond to Nb$^{3+}$ and Nb$^{4+}$, respectively. In the high-temperature phase, each Nb ion has 7/4 electrons in the $4d$ orbitals in average. In the intermediate-temperature phase, since Nb$^{3+}$ and Nb$^{4+}$ ions have two and one electrons in the $4d$ orbitals, the total number of $4d$ electrons per Nb$_4$ cluster is $(2e \times 3) + (1e \times 1) = 7e$.

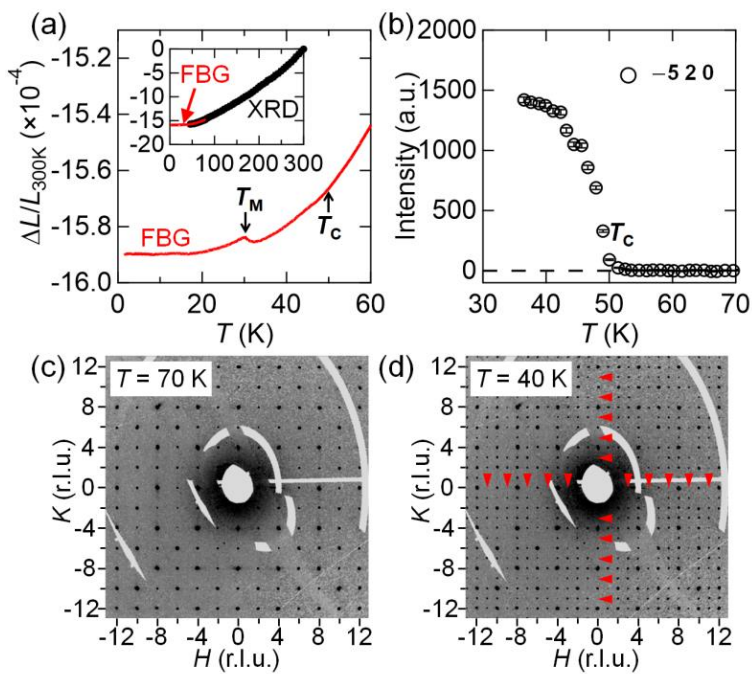

**Figure 3.** (a) Temperature dependence of thermal expansion $\Delta L/L_{300K}$ along the [111] axis of GaNb$_4$Se$_8$. The inset shows the data up to 300 K. Black dots and red lines indicate the XRD and FBG data, respectively. (b) Temperature dependence of the integrated intensity of the $-5\ 2\ 0$ reflection. (c,d) XRD data on the $H\ K\ 0$ plane at (c) 70 K in the high-temperature phase and (d) 40 K in the intermediate-temperature phase. Red triangles in (d) indicate positions $h00$: $h = 2n+1$ and $0k0$: $k = 2n+1$, where no Bragg peaks are observed.

Figure 4c shows the molecular orbital scheme, indicating the Nb cluster rearrangement from a tetramer in the high-temperature phase to a trimer-monomer in the intermediate-temperature phase. The Nb-Nb distance in the regular tetrahedron in the high-temperature phase is 3.0345(4) Å (Figure 1a), whereas the Nb1-Nb1 and Nb1-Nb2 distances are 2.9730(4) and 3.1036(4) Å, respectively, in the distorted tetrahedron in the intermediate-temperature phase (Figure 4d). The changes in Nb-Nb distances are approximately 2%, which is significantly larger than the distortion in the overall lattice with a change of less than 0.0001%. In the distorted tetrahedron, three Nb1 atoms form a regular trimer, where two of the three $t_{2g}$ ($d_{yz}$, $d_{zx}$, and $d_{xy}$, shown by red, green, and blue ribbons, respectively, in Figure 4e) orbitals on each Nb1 site form $\sigma$ bonds. Since a $Nb^{3+}$ ion has two $4d$ electrons, two electrons are accommodated in each $\sigma$-bonding orbital,[41] resulting in the spin-singlet formation. These $\sigma$-bonding orbitals correspond to the lower-lying singlet and doublet with three up and three down electrons (Figure 4c). This type of spin-singlet trimer is reported in layered compounds $LiVO_2$[42] and $LiVS_2$.[43] A localized $S =$ 1/2 spin, represented by an orange arrow in Figure 4c, remains on each Nb2 site in the intermediate-temperature phase. Therefore, each $Nb_4$ cluster hosts $S = 1/2$ in both the high- and intermediate-temperature phases. However, the unpaired electron is tightly localized on the Nb2 site in the intermediate-temperature phase in contrast to the high-temperature phase. As a result, since the interaction between localized spins in the intermediate-temperature phase is weaker than that in the high-temperature phase, the magnetic susceptibility slightly increases below $T_C$ (Figure 2a). The phase transition at $T_C$ does not involve changes in the cubic lattice or the total magnetic moment. The entropy change of $\Delta S_C = 1.0$ J/(mol K) estimated from the specific heat (Figure 2b) may be mostly attributed to a pure charge component. Considering the degeneracy of the $t_2$ orbital with $S = 1/2$ in the high-temperature phase (Figure 1c), the released entropy is estimated as $R \ln 3$. The observed $\Delta S_C = 1.0$ J/(mol K) is significantly smaller than $R \ln 3 \cong 9.1$ J/(mol K), suggesting the growth of short-range interactions in the high-temperature phase.

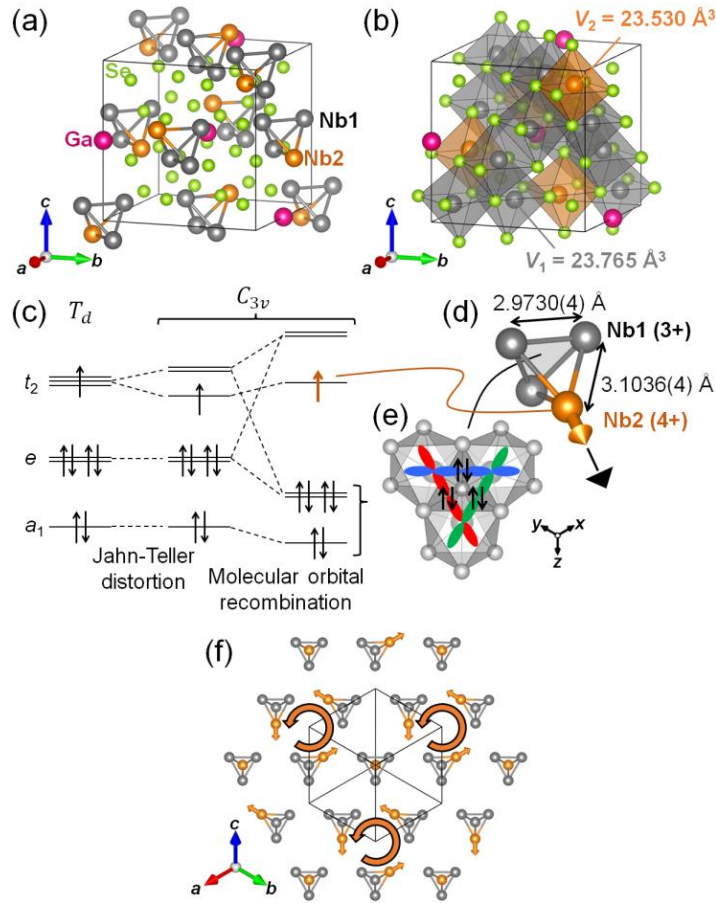

**Figure 4.** Crystal structure of GaNb$_4$Se$_8$ at 40 K in the intermediate-temperature phase, focusing on (a) Nb$_4$ tetrahedra and (b) NbSe$_6$ octahedra. (c) Schematic of the molecular orbital recombination indicating the Nb cluster rearrangement from a tetramer to a trimer-monomer. (d) A distorted Nb$_4$ tetrahedron around 1/2, 0, 1/2 composed of an Nb1 trimer (gray) and an Nb2 atom (orange). A solid triangle indicates the 3-fold rotation axis, along which a local electric dipole moment is present (an orange vector). (e) Schematic of a spin-singlet formation in a Nb1 trimer. Red, green, and blue ribbons indicate the $d_{yz}$, $d_{zx}$, and $d_{xy}$ orbitals extending in the trimer plane, respectively. The $x$, $y$, and $z$ vectors indicate the local axes. (f) Vortex arrangement of electric dipole moments at the tetramers on a (111) layer.

A local electric dipole moment on the distorted Nb tetramer is shown by an orange vector in Figure 4d. The electric dipole moments are oriented in four different directions between adjacent tetramers. When viewing one (111) layer, the electric dipole moments form vortices as shown in Figure 4f, corresponding to a chiral charge order. The structural analysis revealed that two enantiomorphic domains exist in equal proportions (Table S4). These two types of domains correspond to the clockwise and counterclockwise vortices of the electric dipole moments. Figure 4f shows the domain corresponding to the counterclockwise vortices. The two domains may show the optical rotation of opposite signs.

A similar elongation of $M_4$ tetrahedron along one 3-fold rotation axis is observed in GaV$_4$S$_8$[30] and GaV$_4$Se$_8$,[28] where a V ion has 7/4 3$d$ electrons on average. In GaV$_4$S$_8$, all of the tetrahedra are elongated along the same 3-fold rotation axis and the crystal structure changes to polar $R3m$ in the low-temperature phase, where two different V sites form larger and smaller VS$_6$ octahedra with a ratio of 1:3 (Figure S4).[30] If the V site with the larger (smaller) volume of the octahedron is assigned to the V$^{3+}$ (V$^{4+}$) ion with two (one) 3$d$ electrons, considering the ionic radius,[40] the total number of 3$d$ electrons per V$_4$ cluster becomes 5$e$, which differs from the intermediate-temperature phase of GaNb$_4$Se$_8$ with 7$e$ per Nb$_4$ cluster. Therefore, the low-temperature phase in GaV$_4$S$_8$ is not explained by the charge ordering picture but by the simple cooperative Jahn-Teller distortion of the V$_4$ molecular orbitals. In the case of the simple Jahn-Teller scenario, the strain interaction between adjacent $M_4$ clusters is ferroic, resulting in the rhombohedral distortion along the <111> axes like GaV$_4$S$_8$[30] and GaV$_4$Se$_8$.[28] On the other hand, in the case of the charge ordering, the Coulomb interaction favors the antiferroic arrangement between adjacent $M_4$ clusters like $CsW_2O_6$.[13] The structure realized by this antiferroic arrangement corresponds to the chiral charge order with the cubic symmetry. In the dipole-dipole approximation, it should be noted that the Coulomb energy among the distorted tetrahedra is lower for the cubic (antiferroic) case than for the rhombohedral (ferroic) case (Figure S5 in Supporting Information). Considering that the $4d$ orbitals of Nb are more spatially spread than the $3d$ orbitals of V, the orbital overlap between the $M$ sites and the energy gain attributed to the trimer formation (Figure 4c) are larger in $M =$ Nb than in $M =$ V. As a result, the CO phase distinct from the rhombohedral distortion is realized only in $M =$ Nb. The activation energies are reported to be 124, 165, and 49 meV in the high-, intermediate-, and low-temperature phases of $GaNb_4Se_8$ from the electric resistivity.[39] The highest activation energy in the intermediate-temperature phase also supports the charge ordering model.

In the case of $CsW_2O_6$, each $W^{5.33+}$ trimer forms a three-centered-two-electron state and the rest of W ions behave as $W^{6+}$ with empty $5d$ orbitals below the metal-insular transition temperature, resulting in a nonmagnetic insulating state.[13] Therefore, the trimer formation and the nonmagnetic state simultaneously emerge. In $GaNb_4Se_8$, a nonmagnetic phase appears below $T_M = 31$ K, which is lower than the charge ordering temperature $T_C = 50$ K. The low-temperature nonmagnetic state of $GaNb_4Se_8$ may originate from the rearrangement of charges. One may note that the undistorted paramagnetic phase may change directly into nonmagnetic states in $GaNb_4S_8$[32] and $GaTa_4Se_8$[20] due to the absence of the charge order phase.

CONCLUSIONS

The crystal structure in the intermediate-temperature phase of $GaNb_4Se_8$ is investigated by synchrotron XRD using a high-quality single crystal. The Nb cluster rearrangement from tetramer to trimer-monomer is driven by the chiral charge order. The $Nb^{3+}$ trimer is stabilized by $\sigma$-bonding six $4d$ electrons, while a localized $S$ = 1/2 spin remains on the $Nb^{4+}$ ion. The chiral charge order maintains cubic symmetry, which is not found in other lacunar spinel $GaM_4X_8$ series.

ASSOCIATED CONTENT

**Supporting Information.**

The Supporting Information is available free of charge at ACS Publications website. Results of the structural analysis, magnetic susceptibility, and heat capacity (PDF)

**Corresponding Author**

Shunsuke Kitou - Department of Advanced Materials Science, The University of Tokyo, Kashiwa 277-8561, Japan.; E-mail: kitou@edu.k.u-tokyo.ac.jp

**Notes**

The authors declare no competing financial interest.

ACKNOWLEDGMENT

We thank H. Ishikawa and Y. Okamoto for fruitful discussions, and A. Ikeda for generously allowing us to use an optical sensing instrument (Hyperion si155, LUNA) for thermal expansion. This work was supported by Grants-in-Aid for Scientific Research (No. JP19H05826, JP22K14010, JP23K13068, and JP23H01120) from JSPS. The synchrotron radiation experiments were performed at SPring-8 with the approval of the Japan Synchrotron Radiation Research Institute (JASRI) (Proposal No. 2023B1603).

# Supporting Information of

## Cluster rearrangement by chiral charge order in lacunar spinel GaNb$_4$Se$_8$

Shunsuke Kitou[1], Masaki Gen[2], Yuiga Nakamura[3], Yusuke Tokunaga[1], Taka-hisa Arima[1,2]

[1]*Department of Advanced Materials Science, The University of Tokyo, Kashiwa 277-8561, Japan.*

[2]*RIKEN Center for Emergent Matter Science, Wako 351-0198, Japan.*

[3]*Japan Synchrotron Radiation Research Institute (JASRI), SPring-8; Hyogo 679-5198, Japan.*

Table S1. Structural parameters of $GaNb_4Se_8$ at 70 K. The space group is $F\bar{4}3m$ (No. 216) and $a$ = 10.4143(11) Å. Note that $U_{11} = U_{22} = U_{33}$ and $U_{12} = U_{13} = U_{23}$.

| Atom | Wyckoff position | Site symmetry | $x$ | $y$ | $z$ | $U_{11}$ (Å$^2$) | $U_{12}$ (Å$^2$) |
|---|---|---|---|---|---|---|---|
| Ga | 4$a$ | $\bar{4}3m$ | 0 | 0 | 0 | 0.003035(13) | 0 |
| Nb | 16$e$ | .3$m$ | 0.396982(4) | = $x$ | = $x$ | 0.004115(8) | -0.000372(6) |
| Se1 | 16$e$ | .3$m$ | 0.635612(5) | = $x$ | = $x$ | 0.004135(9) | 0.000178(8) |
| Se2 | 16$e$ | .3$m$ | 0.134355(4) | = $x$ | = $x$ | 0.003183(8) | -0.000248(7) |

Table S2. Summary of crystallographic data of $GaNb_4Se_8$ at 70 K.

| | |
|---|---|
| Wavelength (Å) | 0.30956 |
| Crystal dimension ($\mu m^3$) | 52 × 48 × 35 |
| Space group | $F\bar{4}3m$ |
| $a$ (Å) | 10.4143(11) |
| Z | 4 |
| $F$(000) | 1868 |
| $(\sin\theta/\lambda)_{max}$ (Å$^{-1}$) | 1.79 |
| $N_{total}$ | 44710 |
| $N_{unique}$ | 2629 |
| Average redundancy | 17.006 |
| Completeness (%) | 100 |
| Number of unique reflections ($I$>3$\sigma$ / all) | 2575 / 2629 |
| Assuming Ga occupancy = 1 ($N_{parameters}$ = 12) | |
| $R_1$ ($I$>3$\sigma$ / all) | 2.14% / 2.17% |
| $wR_2$ ($I$>3$\sigma$ / all) | 2.46% / 2.47% |
| GOF ($I$>3$\sigma$ / all) | 1.76 / 1.74 |
| Flack parameters ($a,b,c$:-$a$,-$b$,-$c$ ($\bar{1}$)) | 0.994(12) : 0.006(12) |
| CCDC | 2323621 |
| Allowing Ga deficiency ($N_{parameters}$ = 13) | |
| Occupancy of Ga | 0.994(3) |
| $R_1$ ($I$>3$\sigma$ / all) | 2.14% / 2.17% |
| $wR_2$ ($I$>3$\sigma$ / all) | 2.46% / 2.47% |
| GOF ($I$>3$\sigma$ / all) | 1.75 / 1.74 |

Table S3. Structural parameters of GaNb$_4$Se$_8$ at 40 K. The space group is $P2_13$ (No. 198) and $a$ = 10.4138(11) Å.

| Atom | Wyckoff position | Site symmetry | $x$ | $y$ | $z$ |
|------|------|------|------|------|------|
| Ga | 4$a$ | .3. | 0.498942(6) | = $x$ | = $x$ |
| Nb1 | 12$b$ | 1 | 0.598205(4) | 0.895622(3) | 0.602942(4) |
| Nb2 | 4$a$ | .3. | 0.893013(5) | = $x$ | = $x$ |
| Se1 | 12$b$ | 1 | 0.635199(5) | 0.368042(5) | 0.362978(5) |
| Se2 | 12$b$ | 1 | 0.634497(5) | 0.868026(5) | 0.363490(5) |
| Se3 | 4$a$ | .3. | 0.633245(5) | = $x$ | = $x$ |
| Se4 | 4$a$ | .3. | 0.139546(6) | = $x$ | = $x$ |

| Atom | $U_{11}$ (Å$^2$) | $U_{22}$ (Å$^2$) | $U_{33}$ (Å$^2$) | $U_{12}$ (Å$^2$) | $U_{13}$ (Å$^2$) | $U_{23}$ (Å$^2$) |
|------|------|------|------|------|------|------|
| Ga | 0.002251(8) | = $U_{11}$ | = $U_{11}$ | -0.000016(8) | = $U_{12}$ | = $U_{12}$ |
| Nb1 | 0.002112(9) | 0.002401(10) | 0.002292(11) | 0.000014(6) | -0.000032(8) | -0.000403(9) |
| Nb2 | 0.002686(8) | = $U_{11}$ | = $U_{11}$ | -0.000148(7) | = $U_{12}$ | = $U_{12}$ |
| Se1 | 0.002149(12) | 0.002187(10) | 0.002192(13) | 0.000077(8) | 0.000145(8) | -0.000133(9) |
| Se2 | 0.002640(14) | 0.002726(12) | 0.002549(12) | -0.000074(10) | 0.000325(9) | 0.000021(9) |
| Se3 | 0.002193(9) | = $U_{11}$ | = $U_{11}$ | -0.000123(9) | = $U_{12}$ | = $U_{12}$ |
| Se4 | 0.002430(9) | = $U_{11}$ | = $U_{11}$ | 0.000004(9) | = $U_{12}$ | = $U_{12}$ |

Table S4. Summary of crystallographic data of GaNb$_4$Se$_8$ at 40 K.

| | |
|---|---|
| Wavelength (Å) | 0.30956 |
| Crystal dimension ($\mu$m$^3$) | 52 × 48 × 35 |
| Space group | $P2_1 3$ |
| $a$ (Å) | 10.4138(11) |
| Z | 4 |
| $F(000)$ | 1868 |
| $(\sin\theta/\lambda)_{max}$ (Å$^{-1}$) | 1.79 |
| $N_{total}$ | 178930 |
| $N_{unique}$ | 17764 |
| Average redundancy | 17.006 |
| Completeness (%) | 99.56 |
| $N_{parameters}$ | 43 |
| Number of unique reflections ($I$>3$\sigma$ / all) | 17159 / 17764 |
| $R_1$ ($I$>3$\sigma$ / all) | 2.24% / 2.34% |
| $wR_2$ ($I$>3$\sigma$ / all) | 2.67% / 2.69% |
| GOF ($I$>3$\sigma$ / all) | 1.23 / 1.22 |
| Flack parameters ($a,b,c$:-$a$,-$b$,-$c$ ($\bar{1}$):-$b$,-$a$,$c$ (..$m$):$b$,$a$,-$c$ (..2)) | 0.490(15) : 0.008(8) : 0.499(8) : 0.003(8) |
| CCDC | 2323622 |

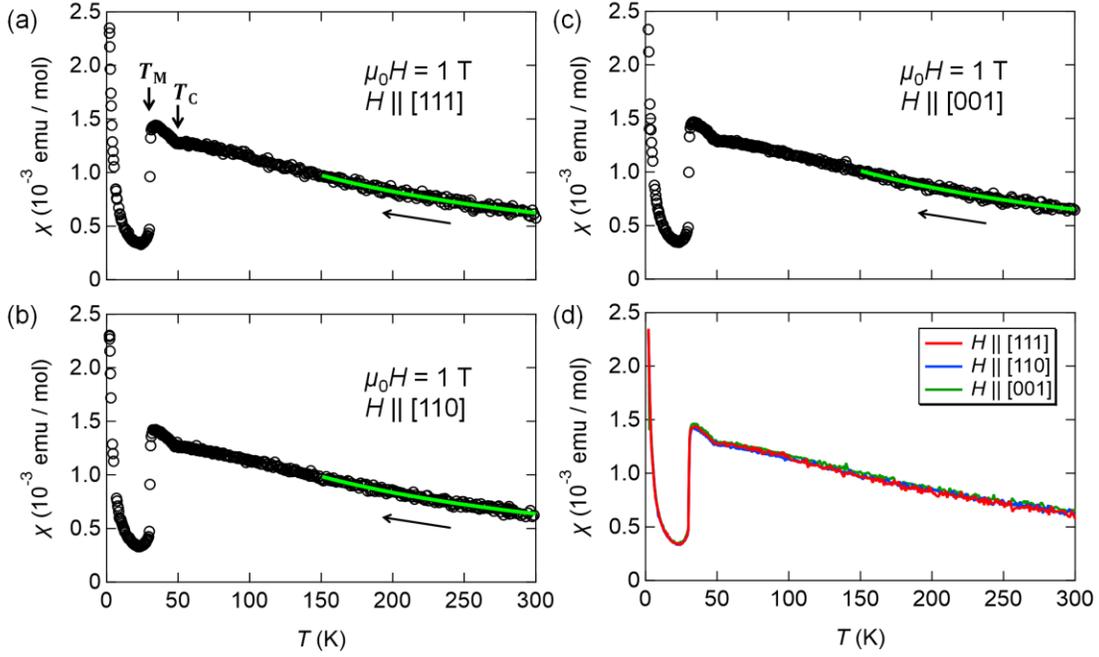

Figure S1. Temperature dependence of the magnetic susceptibility $\chi$ for (a) $H \parallel [111]$, (b) $H \parallel [110]$, and (c) $H \parallel [001]$ of GaNb$_4$Se$_8$. Yellow-green lines indicate the results of Curie-Weiss fits above 150 K. (d) The three sets of data (a)-(c) are displayed overlapping.

Table S5. Summary of Curie-Weiss fits for the magnetic susceptibility using a formula $\chi(T) = \chi_0 + C/(T - \Theta)$, where $\chi_0$, $C$, and $\Theta$ indicate temperature independent terms of $\chi$, Curie constant, and Weiss temperature.

|  | $\chi_0$ [emu/mol] | $C$ [K emu/mol] | $\Theta$ [K] | $\mu_{\text{eff}}$ [$\mu_B$/f. u.] |
|---|---|---|---|---|
| $H \parallel [111]$ | $-6.2 \times 10^{-5}$ | 0.3033 | $-142.6$ | 1.558 |
| $H \parallel [110]$ | $-4.6 \times 10^{-5}$ | 0.3024 | $-143.1$ | 1.555 |
| $H \parallel [001]$ | $-4.7 \times 10^{-5}$ | 0.3063 | $-139.2$ | 1.565 |

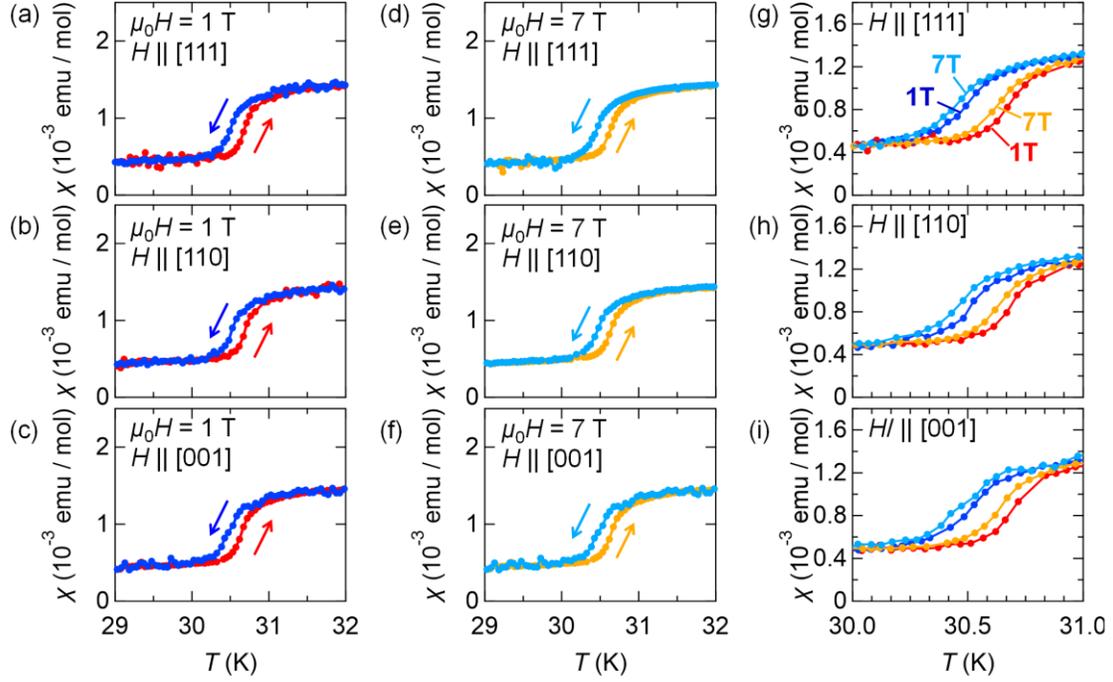

Figure S2. Enlarged views of temperature dependence of the magnetic susceptibility $\chi$ for (a) $H \parallel [111]$, (b) $H \parallel [110]$, and (c) $H \parallel [001]$ under the application of $\mu_0 H = 1$ T, and for (d) $H \parallel [111]$ (e) $H \parallel [110]$, and (f) $H \parallel [001]$ under the application of $\mu_0 H = 7$ T. (g)-(i) The data for different external magnetic fields are displayed overlapping. The entropy change at $T_M$ is estimated from the temperature dependence and the external magnetic field dependence of the magnetic susceptibilities using the Clausius-Clapeyron equation $\Delta S_M = -\Delta M \cdot dB/dT_M$. $\Delta M$ is calculated using the average difference in the magnetization before and after the magnetic phase transition at $\mu_0 H = 1$ T and 7 T. Since $\Delta M = 0.034$ J/mol T and $dB/dT_M = 6$ T/$-0.03$ K are estimated from $\chi$ for $H \parallel [111]$ under $\mu_0 H = 1$ T and 7 T, the magnetic entropy change is calculated as $\Delta S_M = 6.8$ J/mol K, which is comparable to the total magnetic entropy $R \ln 2 \cong 5.76$ J/mol K expected for $S_{\text{eff}} = 1/2$.

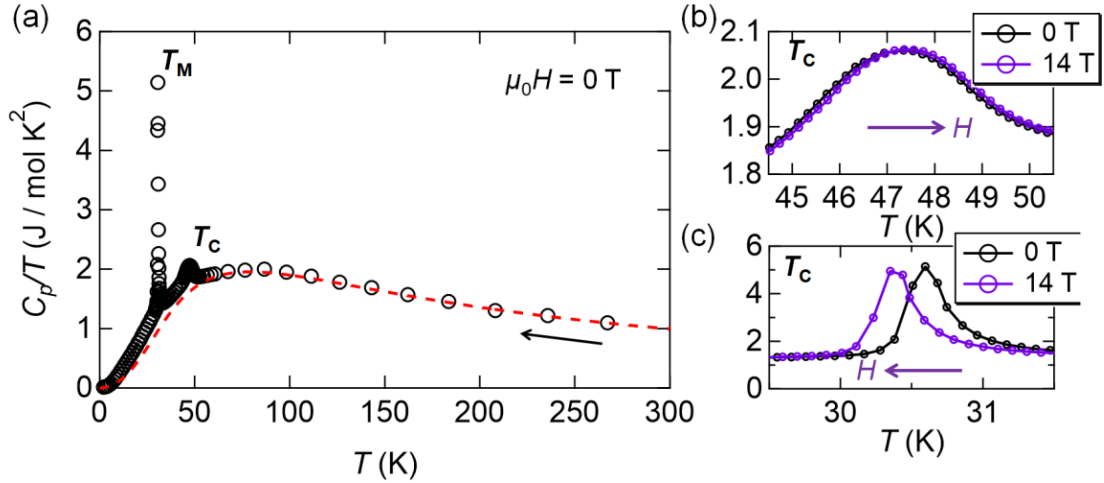

Figure S3. (a) Temperature dependence of the heat capacity divided by temperature $C_p/T$ at 0 T. The data were obtained in a cooling run. The red dashed line denotes the estimated lattice heat capacity $C_{\mathrm{fit}}$ based on a two-phonon Debye model. Here, the fitting functions are defined as $C_{\mathrm{fit}} = \frac{1}{3}C_{\mathrm{D_1}} + \frac{2}{3}C_{\mathrm{D_2}}$, and $C_{\mathrm{D}_i} = 9Nk_B\left(\frac{T}{\Theta_{D_i}}\right)^3\int_0^{\frac{\Theta_{D_i}}{T}}\frac{x^4e^x}{(e^x-1)^2}\,dx$, where $N = 13$ is the number of atoms in the formula unit. The obtained fitting parameters are $\Theta_{\mathrm{D1}} = 600$ K and $\Theta_{\mathrm{D2}} = 250$ K. As mentioned in the main manuscript, the Debye model fails to reproduce the experimental data because a $T$-bilinear component appears at low temperatures. (b,c) Enlarged views around (b) $T_{\mathrm{C}}$ and (c) $T_{\mathrm{M}}$. Black and purple dots indicate $C_p/T$ for $H = 0$ and for $\mu_0H = 14$ T along [111], respectively. $T_{\mathrm{M}}$ decreases with increasing the external magnetic fields, while $T_{\mathrm{C}}$ slightly increases.

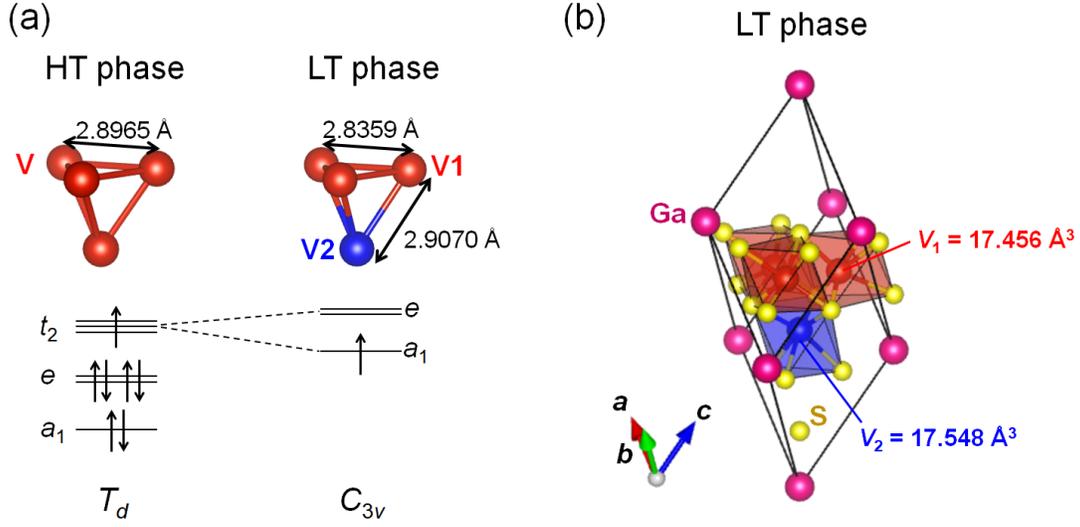

Figure S4. Schematic of a cooperative Jahn-Teller distortion of the $t_2$ molecular orbitals and crystal structure of GaV$_4$S$_8$ at 20 K in the low-temperature phase, referred from [1]. The volumes of V1S$_6$ and V2S$_6$ octahedra are 17.456 and 17.548 Å$^3$, respectively, which contradicts the model of V$^{3+}$ trimer-V$^{4+}$ monomer formation.

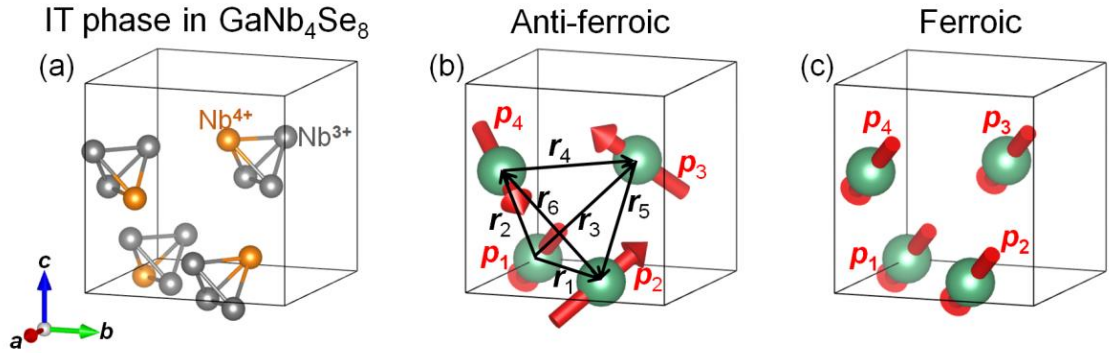

Figure S5. The estimation of Coulomb interactions using the dipole-dipole approximation. (a) The arrangement of the distorted Nb tetramer in the intermediate-temperature (IT) phase of GaNb$_4$Se$_8$. (b) The anti-ferroic arrangement of electric dipole moments (red vectors) corresponding to the IT phase with cubic symmetry. (c) The ferroic arrangement of electric dipole moments with rhombohedral symmetry, similar to GaV$_4$S$_8$ [1]. The dipole-dipole interaction is represented by $V_{ij} = \frac{1}{4\pi\varepsilon_0 r^3}\left[\boldsymbol{p}_i \cdot \boldsymbol{p}_j - \frac{3(\boldsymbol{p}_i \cdot \boldsymbol{r})(\boldsymbol{p}_j \cdot \boldsymbol{r})}{r^2}\right]$. Here, $\boldsymbol{p}_i$ and $\boldsymbol{p}_j$ are the electric dipole moments, and $\boldsymbol{r}$ is the vector connecting two electric dipoles. In the case of the anti-ferroic arrangement, the Coulomb interaction between each dipole in the four tetramers is $V = \frac{1}{4\pi\varepsilon_0 r^3}\left(-\frac{p^2}{3}\right)$. In the case of the ferroic arrangement, the averaged Coulomb interaction between each dipole in the four tetramers is $V = 0$. Therefore, the Coulomb energy among the distorted tetrahedra is lower in the anti-ferroic arrangement than in the ferroic arrangement, resulting in the emergence of chiral charge order in GaNb$_4$Se$_8$.